\documentclass[preprint,10pt]{elsarticle}
\usepackage[margin=1in,heightrounded]{geometry}
\usepackage{hyperref}
\usepackage{amsmath, amsthm, amsfonts, amssymb}
\usepackage{algorithm}
\usepackage{algorithmic}
\usepackage{subcaption}

\newtheorem{theorem}{Theorem}
\newtheorem{lemma}{Lemma}
\newtheorem{corollary}{Corollary}

\newtheorem{definition}{Definition}
\newtheorem{remark}{Remark}

\begin{document}
\begin{frontmatter}
\title{A Fast Algorithm for Source-wise Round-trip Spanners}


\author[uncg]{Chun Jiang Zhu\corref{cor1}}
\ead{chunjiang.zhu@uncg.edu}

\author[uconn]{Song Han}
\ead{song.han@uconn.edu}

\author[cityu]{Kam-Yiu~Lam}
\ead{cskylam@cityu.edu.hk}

\address[uncg]{Dept. of Computer Science, Univ. of North Carolina at Greensboro, Greensboro, NC, USA}

\address[uconn]{Dept. of Computer Science and Engineering, University of Connecticut, Storrs, CT, USA}

\address[cityu]{Dept. of Computer Science, City University of Hong Kong, Hong Kong, China}

\cortext[cor1]{Corresponding author. Address: Department of Computer Science, University of North Carolina at Greensboro,
P.O. Box 26170, Greensboro, NC 27402-6170}


\begin{abstract}
In this paper, we study the problem of fast constructions of source-wise round-trip spanners in weighted directed graphs.
For a source vertex set $S\subseteq V$ in a graph $G(V,E)$, an $S$-sourcewise round-trip spanner of $G$ of stretch $k$ is a subgraph $H$ of $G$ such that for every pair of vertices $u,v\in S\times V$, their round-trip distance in $H$ is at most $k$ times of their round-trip distance in $G$.
We show that for a graph $G(V,E)$ with $n$ vertices and $m$ edges, an $s$-sized source vertex set $S\subseteq V$ and an integer $k>1$, there exists an algorithm that in time $O(ms^{1/k}\log^5n)$ constructs an $S$-sourcewise round-trip spanner of stretch $O(k\log n)$ and $O(ns^{1/k}\log^2n)$ edges with high probability.
Compared to the fast algorithms for constructing all-pairs round-trip spanners \cite{PRS+18,CLR+20}, our algorithm improve the running time and the number of edges in the spanner when $k$ is super-constant. Compared with the existing algorithm for constructing source-wise round-trip spanners \cite{ZL17}, our algorithm significantly improves their construction time $\Omega(\min\{ms,n^\omega\})$ (where $\omega \in [2,2.373)$ and 2.373 is the matrix multiplication exponent) to nearly linear $O(ms^{1/k}\log^5n)$, at the expense of paying an extra $O(\log n)$ in the stretch.
As an important building block of the algorithm, we develop a graph partitioning algorithm to partition $G$ into clusters of bounded radius and prove that for every $u,v\in S\times V$ at small round-trip distance, the probability of separating them in different clusters is small.
The algorithm takes the size of $S$ as input and does not need the knowledge of $S$.
With the algorithm and a reachability vertex size estimation algorithm, we show that the recursive algorithm for constructing standard round-trip spanners \cite{PRS+18} can be adapted to the source-wise setting.
We rigorously prove the correctness and computational complexity of the adapted algorithms.
Finally, we show how to remove the dependence on the edge weight in the source-wise case.
\end{abstract}

\end{frontmatter}
\setcounter{section}{0}

\section{Introduction}
Graph spanners are sparse graph structures approximating shortest path distances in graphs and have received considerable research interests since they were proposed in the late 80's \cite{PS89}.
A spanner of \emph{stretch} $\alpha$ ($\alpha$-spanner) for an undirected graph $G(V,E)$ is a subgraph $H(V,E'\subseteq E)$ of $G$ such that for every pair of vertices $u,v\in V$, their distance in $H$ is at most $\alpha$ times their original distance in $G$.
It is well-known that for an integer $k>1$, every graph on $n$ vertices has a $(2k-1)$-spanner of size (number of edges) $O(n^{1+1/k})$ \cite{ADD+93,TZ05}.
The stretch-size trade-off is optimal if we believe the Erdos's girth Conjecture \cite{Erdos64}.
Research efforts were then devoted to $\beta$-additive spanners and $(\alpha,\beta)$-spanners.
In the formers, the distance between every vertex pair is no larger than their distance in the original graph by only an additive term $\beta$.
Existing studies show that for an undirected graph, there exists a $2$-additive spanner of size $O(n^{3/2}\log n)$ \cite{ACI+99} (with the log factor shaved in \cite{EP04}), a $4$-additive spanner of size $\widetilde{O}(n^{7/5})$ \cite{Chechik13}, and a 6-additive spanner of size $O(n^{4/3})$ \cite{BKM+10,Woodruff10}, where the notation $\widetilde{O}(\cdot)$ hides polylogarithmic factors.
Abboud \emph{et al.} \cite{AB16} proved that one cannot obtain an additive spanner with a constant surplus $\beta=O(1)$ using size $O(n^{4/3-\epsilon})$.
In the latter, the distance between every vertex pair is no larger than $\alpha\cdot d+\beta$ where $d$ is their distance in the original graph. For the different constructions of $(\alpha,\beta)$-spanners and other variants of spanners, readers are referred to \cite{EP04,TZ06,Pettie09,BKM+10,ADF+19,ABD+20,ABS+20,BDR21}.

The definition of spanners can be naturally extended to directed graphs (digraphs) but it becomes trivial to study spanners in this setting because of the well-known lower bound $\Omega(n^2)$ on the size \cite{RTZ08}.
Instead of studying one-way distances in digraphs, Cowen and Wagner \cite{CW99,CW00} studied round-trip distances for the first time.
The \emph{round-trip distance} between vertex $u$ and $v$ in a digraph $G$ is the sum of the one-way distance from $u$ to $v$ and the one-way distance from $v$ to $u$ in $G$.
A \emph{$k$-roundtrip spanner} of $G$ is a subgraph $H$ of $G$ such that for every $u,v$ in $G$, their round-trip distance in $H$ is at most $k$ times of their original round-trip distance in $G$.
In \cite{CW99,CW00}, Cowen and Wagner studied the round-trip routing schemes which implied a $(2^k-1)$-roundtrip spanner of size $\widetilde{O}(n^{1+1/k})$.
Later, Roditty \emph{et al.} \cite{RTZ08} significantly improved the spanner stretch by proposing a randomized algorithm that constructs a $(2k+\epsilon)$-roundtrip spanner of size $O(\min\{(k^2/\epsilon)n^{1+1/k}\log(nw),(k/\epsilon)^2n^{1+1/k}\log^{2-1/k}n\})$ essentially in time $\Omega(\min\{mn,n^\omega\})$ with $\omega \in [2,2.373)$ (2.373 is the matrix multiplication exponent), for a graph with $n$ vertices, $m$ edges, maximum edge weight $w$, and a parameter $\epsilon>0$.
Zhu and Lam \cite{ZL18} developed a deterministic algorithm that constructs a $(2k+\epsilon)$-roundtrip spanner of size $O((k/\epsilon)n^{1+1/k}\log(nw))$ in the same running time, reducing the size of the spanner of Roditty \emph{et al.} \cite{RTZ08} by a factor of $k$.
Recently, Cen \emph{et al.} \cite{CDG20} further reduced the stretch to $2k-1$. Their deterministic algorithm constructs a $(2k-1)$-roundtrip spanner of size $O(kn^{1+1/k}\log n)$ in $\widetilde{O}(kmn\log w)$ time.

However, the runtimes of the algorithms \cite{RTZ08,ZL18,CDG20} are $\Omega(\min\{mn,n^\omega\})$, which is essentially the same as that of computing All-Pairs Shortest Paths (APSP) \cite{Orlin17,WW10}.
Pachocki \emph{et al.} \cite{PRS+18} proposed a nearly linear time $O(mn^{1/k}\log^5n)$ algorithm that constructs an $O(k\log n)$-roundtrip spanner of size $O(n^{1+1/k}\log^2n)$ with high probability.
Their algorithm combines exponential distribution based graph decomposition \cite{MPX13} and sampling based reachability vertex size estimation \cite{Cohen97}.
The former enables reasoning about the probability of separating two vertices of bounded round-trip distance in different clusters.
In the graph decomposition, they choose a subset of vertices as the centering vertices to construct clusters such that their algorithm terminates in only logarithmic levels of recursions.
They also proposed an algorithm that constructs a nearly tight additive round-trip spanner of additive term $O(n^\alpha)$ and size $\widetilde{O}(n^{2-\alpha})$ in time $\widetilde{O}(mn^{1-\alpha})$ with high probability, for an unweighted graph and a parameter $\alpha \in (0,1)$.
Recently, Chechik et al. \cite{CLR+20} improved the multiplicative spanner stretch to $O(k\log\log n)$ and $O(k\log k)$ respectively, while using the same run-time as the algorithm in \cite{PRS+18}. In addition, they developed an $\tilde{O}(m\sqrt{n})$ time algorithm that computes an $(8+\epsilon)$-roundtrip spanner of expected $O(n^{1.5})$ edges. Later in \cite{DW20}, the stretch factor was improved to $(5+\epsilon)$ for the same running time.

{\noindent \bf Source-wise Spanners.}
Another research thread focuses on constructing graph structures that approximate distances only for \emph{some pairs} of vertices, instead of all pairs.
Coppersmith and Elkin \cite{CE06} for the first time studied exact pair-wise preservers.
Given an undirected graph $G(V,E)$ and a set of pairs of vertices, $P \subseteq V \times V$, a subset of $G$, $H$, is called a $P$-pairwise preserver of $G$ if for every vertex pair in $P$, their distance in $H$ is the same as that in $G$.
They proved that for a set of vertex pairs $P$ of size $p$, every graph $G$ contains a $P$-pairwise preserver of size $O(\min\{np^{1/2},n+n^{1/2}p\})$.
Recently, Bodwin \cite{Bodwin17} obtained a size bound of the pair-wise preserver, $O(n+n^{2/3}p)$, even in the case of digraphs.
Roditty \emph{et al.} \cite{RTZ05} worked on source-wise spanners. For a set of \emph{sources} $S\subseteq V$, a subset of $G$, $H$, is called an $S$-sourcewise $k$-spanner of $G$ if for every vertex pair in $S\times V$, their distance in $H$ is at most $k$ times of the distance in $G$.
They proposed an algorithm that constructs an $S$-sourcewise $(2k-1)$-spanner of size $O(kns^{1/k})$ in expected $\widetilde{O}(kms^{1/k})$ time, where $s$ is the size of $S$.
Further work along this direction studied source-wise spanners with an \emph{additive} stretch.
There have also been polynomial-time algorithms constructing $S$-sourcewise spanners of additive stretches $2$,$4$, and $6$, and sizes $\widetilde{O}(n^{5/4}s^{1/4})$ \cite{KV13}, $\widetilde{O}(n^{11/9}s^{2/9})$, and $O(n^{6/5}s^{1/5})$ \cite{Kavitha15}, respectively.
See other stretch-size trade-offs of source-wise spanners and source-wise spanners for the setting $S\times S$ in \cite{CGK13,Parter14,Pettie09}.

Recently, source-wise round-trip spanners, a natural extension of standard round-trip spanners in the source-wise setting, were firstly studied by Zhu and Lam \cite{ZL17}.
Given a set of sources $S$ in a digraph $G$, a subgraph $H$ of $G$ is called an \emph{$S$-sourcewise $k$-roundtrip spanner} of $G$ if for every $u\in S,v\in V$, their round-trip distance in $H$ is at most $k$ times of their round-trip distance in $G$.
Source-wise round-trip spanners can have applications for computing the source-wise variants of round-trip compact routing schemes, distance oracles and low distortion embeddings \cite{ZL17}.
They proposed an algorithm that in time $\Omega(\min\{ms,n^\omega\})$ constructs an $S$-sourcewise $(2k+\epsilon)$-roundtrip spanner of size $O((k^2/\epsilon)ns^{1/k}\log(nw))$.
However, the construction time of the algorithm, $\Omega(\min\{ms,n^\omega\})$, is quite large, especially when the size $s$ is a large integer.
It is natural to ask whether the fast algorithm for constructing standard round-trip spanners \cite{PRS+18} can be adapted to the source-wise setting.

{\noindent \bf Contributions.}
In this paper, we provide an affirmative answer to the above question and show a generalization of the algorithms of \cite{PRS+18} for the all-pairs construction to the source-wise setting, trading off computation time for stretch.  As an important building block, we first develop a graph partitioning algorithm to partition a graph $G$ into clusters of bounded radius, and prove that for vertex $u$ in a source vertex set $S$ and $v\in V$ at small round-trip distance, the probability of separating them in different clusters is small. The algorithm does not require the knowledge of $S$ but only its size $|S|$. The second algorithm is an algorithm for estimating the sizes of in-balls and out-balls around some center vertices with a radius. Informally, an in-ball (out-ball) around a vertex $u$ with a radius $r$ is the vertices that $u$ can be reachable ($u$ can reach, respectively) within distance $r$. It can perform size estimations for some subset of vertices $U$, instead of all vertices $V$. For a small $U$, the algorithm is more efficient compared to Cohen’s algorithm \cite{Cohen97}. With the two algorithms at hand, we show that the recursive algorithm for constructing standard round-trip spanners \cite{PRS+18} can be adapted to the source-wise setting by thoroughly setting the source vertex set in every recursion. Furthermore, we rigorously prove the correctness and computational complexity of the adapted algorithms. Finally, we show how to remove the dependence of the edge weights in the source-wise setting. Our main result is summarized in the following theorem.

\begin{theorem}[{\bf Improved Source-wise Round-trip Spanners}]
\label{thm:rts}
For a weighted directed graph $G(V,E)$ on $n$ vertices and $m$ edges, a source vertex set $S\subseteq V$ of size $s$ and an integer $k>1$, there is an algorithm that in time $O(ms^{1/k}\log^5n)$ constructs an $S$-sourcewise $O(k\log n)$-roundtrip spanner of size $O(ns^{1/k}\log^2n)$ with high probability.
\end{theorem}

Compared with the existing algorithm \cite{ZL17} for constructing source-wise round-trip spanners, our result significantly improves their construction time $\Omega(\min\{ms, n^\omega\})$ (where $\omega \in [2,2.373)$) to a nearly linear time $O(ms^{1/k}\log^5n)$.
The expense is that the stretch of the spanner is worsened by a factor of $\log n$.
Note that the round-trip spanners preserving alll-pairs round-trip distances are feasible source-wise round-trip spanners.
Compared to the fast algorithm for constructing all-pairs round-trip spanners \cite{PRS+18}, we reduce the size of the spanner from $O(n\cdot n^{1/k}\log^2n)$ to $O(n\cdot s^{1/k}\log^2n)$, and improve the construction time from $O(m\cdot n^{1/k}\log^5n)$ to $O(m\cdot s^{1/k}\log^5n)$ when $k$ is super-constant. 

\begin{remark}
The maximum value of $k$ to consider in the setting of multiplicative spanners is bounded by $O(\log n)$. When $k=O(\log n)$, the performance of the spanner constructed in Theorem \ref{thm:rts} is the same as that in the algorithm of \cite{PRS+18}. When $k$ is a constant, the spanner in Theorem \ref{thm:rts} has stretch $O(\log n)$ and size $\widetilde{O}(n^{1+c})$ for some constant $c$. The size is larger than the nearly linear $\widetilde{O}(n)$ size of the spanner of the same stretch by the algorithm of \cite{CDG20} while the nearly linear run-time is much faster than the $\Omega(mn)$ time in \cite{CDG20}.  However, when $k$ is super-constant, our algorithm can still bring subpolynomial benefits in both the run-time and the size of the spanner compared to the algorithms of \cite{PRS+18,CLR+20}.
\end{remark}


The remainder of this paper is organized as follows.
In Section \ref{sec:notations}, we present the notations and definitions we will use.
Next, we provide technical details of our algorithms for constructing source-wise round-trip covers in Section \ref{sec:rtspanner}, and then describe the removal of the dependence of edge weights in Section \ref{sec:dependence}.
Finally, we conclude the paper with a brief discussion on the future work in Section \ref{sec:conclusion}.


\section{Notations and Definitions} \label{sec:notations}
We consider weighted directed graphs $G(V,E,W)$, where $V$ and $E$ are the vertex set and edge set and $W$ assigns weight $W(e)$ to an edge $e\in E$.  Throughout the paper let $n=|V|$ and $m=|E|$ denote the number of vertices and the number of edges in $G$, and $w=\max_{e\in E}W(e)$ denote the maximum edge weight. Let $G(U)$ denote the subgraph induced by an arbitrary set of vertices $U$. We use $d_G(u,v)$ to denote the \emph{(one-way shortest) distance} from $u$ to $v$ in $G$. The \emph{round-trip distance} between $u$ and $v$ in $G$, denoted by $d_G(u\rightleftarrows v)=d_G(u,v)+d_G(v,u)$, is the distance from $u$ to $v$ plus the distance from $v$ to $u$. We drop the subscript $G$ if it is clear from context. The \emph{radius} and \emph{diameter} of $G$ are defined as $\min_{u\in V} \max_{v\in V} d(u,v)$ and $\max_{u,v\in V} d(u,v)$, respectively. Similarly, the \emph{round-trip radius} of $G$ is $\min_{u\in V} \max_{v\in V} d(u\rightleftarrows v)$. We allow that there may be multiple round-trip shortest paths between two vertices.

A \emph{(round-trip) ball} with round-trip radius $R$ around vertex $u$ in $G(U)$, denoted as $ball_{U}(u, R)$, is the set of vertices with round-trip distance no larger than $R$ from $u$ in $G(U)$. Similarly, let $out$-$ball_{U}(u, r)$ and $in$-$ball_{U}(u, r)$ denote the set of vertices that can be reached from $u$ or can reach $u$ within distance $r$ in $G(U)$, respectively. A shortest path out-tree from vertex $v$ to a subset of vertices $U$ in $G$ is a subgraph $T$ of $G$ s.t. $T$ is a tree rooted at $v$ with all edges oriented away from $v$, and for every $u\in U$, $d_G(v,u)=d_T(v,u)$. Similarly a shortest path in-tree from $U$ to $v$ is a subgraph $T$ of $G$ s.t. $T$ is a tree rooted at $v$ with all edges oriented towards from $v$, and for every $u\in U$, $d_G(u,v)=d_T(u,v)$. We define the \emph{round-trip tree} of a ball $B=ball_{U}(v,R)$, denoted by $RT$-$Tree(B)$, as the union of a shortest path out-tree from $v$ to all vertices of $B$ and a shortest path in-tree from all vertices of $B$ to $v$ in $G(U)$. We say that an event happens \emph{with high probability (w.h.p.)} if it happens with probability $O(1-1/n^c)$, where $n$ is the size of the input of the problem and $c$ is a constant. 


\section{An Algorithm for Constructing Source-wise Round-trip Spanners}
\label{sec:rtspanner}

Following existing methods for constructing round-trip spanners \cite{RTZ08,ZL17,PRS+18}, the key subroutine is the construction of \emph{round-trip covers}:  a collection of round-trip balls of a bounded radius such that each pair of vertices at a small round-trip distance is contained in at least one of the balls. In addition, every vertex is required to be included in a bounded number of balls.
The notion of source-wise round-trip covers is formally defined below, while our result for their efficient construction are summarized in Theorem \ref{thm:rtcover}.

\begin{definition}[Source-wise Round-trip Covers \cite{ZL17}]
For a graph $G(V,E)$ and a vertex set $S\subseteq V$, a collection $C$ of round-trip balls is called an $S$-sourcewise $(k, R)$-cover if and only if (1) each ball in $C$ has round-trip radius at most $k\cdot R$; and (2) for every $u\in S,v\in V$ of round-trip distance at most $R$, they are contained in the same ball $B\in C$.
\end{definition}

\begin{theorem}[{\bf Improved Source-wise Round-trip Covers}]
\label{thm:rtcover}
For a graph $G(V,E)$, a source vertex set $S\subseteq V$ of size $s$, an integer $k>1$ and a parameter $R>0$, there is an algorithm that in time $O(ms^{1/k}\log^4n)$ constructs an $S$-sourcewise $(O(k\log n),R)$-cover $C$ w.h.p.
Furthermore, each vertex $v\in V$ is contained in $O(s^{1/k}\log n)$ balls.
\end{theorem}

Compared to the previous construction of source-wise round-trip covers \cite{ZL17}, the running time is improved to a nearly linear time. But the stretch-size trade-off is worse as there is a $\log n$ factor in the stretch.
With the algorithm for constructing source-wise round-trip covers, we immediately get an algorithm for constructing source-wise round-trip spanners through a standard technique \cite{RTZ08}. Specifically, for each logarithmic scale of round-trip distances $[2^{i-1},2^i)$ where $1\leq i\leq \log_2(2nw)$, we construct source-wise $(O(k\log n),2^i)$-cover $C_i$ w.r.t. $S$. We then take the union of $RT$-$Tree(B)$ for each $B\in \cup_{1\leq i\leq \log_2(2nw)}C_i$ as the resulting $S$-sourcewise $O(k\log n)$-roundtrip spanner. Therefore, the number of edges in the source-wise round-trip spanner and the running time contain a term of $\log(nw)$ as in Theorem \ref{thm:weightrts}.
In the remainder of this section, we first present the fast algorithm for constructing source-wise round-trip covers. Then in the next section, we prove our main theorem (Theorem \ref{thm:rts}) by removing the dependence on the maximum edge weight $w$.

\begin{theorem}[{\bf An intermediate result for source-wise round-trip spanners}]
\label{thm:weightrts}
For a graph $G(V,E)$, a source vertex set $S\subseteq V$ of size $s$ and an integer $k>1$, there is an algorithm that in time $O(ms^{1/k}\log^4 n\cdot \log(nw))$ constructs an $S$-sourcewise $O(k\log n)$-roundtrip spanner of size $O(ns^{1/k}\log n\cdot \log(nw))$ w.h.p.
\end{theorem}

\begin{figure}[t]
     \center
     \begin{subfigure}[t]{.4\linewidth}
        \centering\includegraphics[width=.65\linewidth]{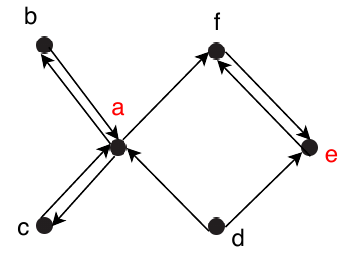}
        \caption{\small The first case}
        \label{fig:casea}
     \end{subfigure}
     \begin{subfigure}[t]{.4\linewidth}
        \centering\includegraphics[width=.55\linewidth]{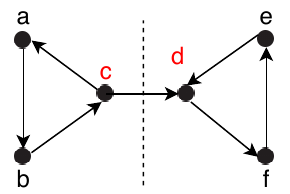}
        \caption{\small The second case}
        \label{fig:caseb}
     \end{subfigure}
     \caption{\small Recursions under the two cases, where source vertices are labeled in red color. (a) There is a vertex $a$ that can reach $\{a,b,c,f\}$ and be reachable from $\{a,b,c,d\}$ within within distance 1. The sizes of both reachability sets are no smaller than a pre-defined threshold 4. Then we include the ball $ball(a,2\cdot 1)=\{a,b,c\}$ into the cover and recurse on the graph induced by the remaining vertices $\{d,e,f\}$. (b) There is no vertex that can reach and be reachable from at least 4 vertices within distance 1. Then we apply a graph partitioning algorithm to get two clusters (separated by the dash line), $\{a,b,c\}$ and $\{d,e,f\}$, and recurse on the graph induced by each cluster.}
     \label{fig:case}
\end{figure}

{\noindent \bf High-Level Ideas.}
Our algorithm is an extension of the fast algorithm for constructing standard round-trip covers \cite{PRS+18} to the source-wise setting. We follow their recursive idea that is based on the following case analysis. If there is a vertex $u$ with a large in- and out-ball of radius $r>0$, then we include the round-trip ball $ball(u, O(r))$ into the cover and recurse on the graph induced by the remaining vertices $G(V-ball(u, O(r)))$. Otherwise, we partition the graph using a source-wise graph partitioning algorithm and then recurse on each part of the partitioning. See the illustrating examples for the two cases in Figure \ref{fig:case}. For both cases, we show that we need to recurse at most a logarithmic number of times because the graph size is reduced by at least a constant factor in each recursion. We are able to prove that every pair $u,v \in S \times V$ at small round-trip distance has a small probability to appear in different balls. Finally, repeating this procedure multiple times and taking union of the obtained balls result in the source-wise round-trip cover.

In order to develop the above algorithm, we need to generalize its component algorithms from the standard setting to the source-wise setting. Specifically, a source-wise graph partitioning algorithm is used to compute a partitioning of vertices in $G$ with the following properties: (i) all parts, except for one, have a bounded round-trip radius, (ii) the last part has bounded size, and (iii) every pair $u,v \in S \times V$ has a small probability to appear in different parts. This algorithm does not require the knowledge of $S$ and only its cardinality $|S|$ suffices to guarantee the desirable properties. A sampling-based algorithm is used to estimate the sizes of in- and out-balls. The recursive algorithm for constructing source-wise round-trip covers is an adaptation of the standard algorithm by adapting the exploration radius and carefully setting the source vertex set in every recursion.

In the sequel, we first present the two subroutines in Sections \ref{subsec:decomp} and \ref{subsec:est}, respectively, and then discuss the algorithm for constructing source-wise round-trip covers in Section \ref{subsec:cover}.

\subsection{A Source-wise Graph Partitioning Algorithm} \label{subsec:decomp}

In order to apply the recursive idea, we need a graph partitioning/clustering algorithm that partitions the vertices of an input graph into clusters of small round-trip radius. More importantly, we need to reason that the probability that the cycle of interest cut by the partitioning (with vertices in different parts) is small. We show that this can be achieved by growing balls with radius deliberately chosen from an exponential distribution with the appropriate parameter. Graph partitioning using the exponential distribution is not new and appeared in several prior works \cite{MPX13,EMP+16,PRS+18}. Miller \emph{et al.} \cite{MPX13} proposed a parallel low-diameter decomposition algorithm on undirected graphs that can reason that not too many edges are cut by the decomposition. Ene \emph{et al.} \cite{EMP+16} studied partitioning for directed graphs instead of undirected graphs. They show that the probability that the partitioning cuts edges cannot be bounded but one can reason the probability that cycles are cut is small. Later, Pachocki \emph{et al.} \cite{PRS+18} enhanced the algorithms of \cite{MPX13,EMP+16} by providing the capability of growing clusters rooted at a specific set of vertices. We take a step further by generalizing the partitioning algorithm in \cite{PRS+18} to the multiple-source setting. For any sources $S\subseteq V$. we can reason that every pair $u,v$ in $S \times V$ at small round-trip distance has a small probability to appear in different parts. Moreover, we observe that the knowledge of the set $S$ is not required in this algorithm.

The partitioning algorithms Cluster-Out and Cluster-In are presented in Algorithm \ref{alg:partitioning}. The input set of vertices $U$ is the chosen set of centers to build clusters while $s$ is the size of a source vertex set $S$.
Cluster-Out independently assigns each vertex $u\in U$ a random radius $r_u$ drawn from the exponential distribution with the parameter $\log (s)/r$. The exponential distribution Exp$(\beta)$ with parameter $\beta$  has probability density function $f(x)=\beta exp(-\beta x)$ on $x\geq 0$.
Note that the parameter we use is $\log (s)/r$ and this is essentially the only crucial change to generalize the algorithm of \cite{PRS+18} to the source-wise setting.
Each vertex $v\in V$ is assigned to the cluster centered at vertex $u\in U$ which maximizes $r_u-d(u,v)$ if the quantity is larger than 0. Otherwise, $v$ is assigned to the last cluster.
The algorithm can be implemented as parallel ball growing similar to \cite{MPX13,EMP+16}.
We first pick a start time $r_u$ for every vertex $u\in U$ from Exp$(\log (s)/r)$, and then explore the graph by starting a search from $u$ at time $r_u$ if $u$ is not already reached by some other vertex. The search takes $W(e)$ time to propagate across an edge $e$, and each time step can be performed in parallel over all vertices in $U$. Each vertex $v$ in $V$ is assigned to the cluster rooted at the vertex $u$ that reaches it first.

The properties of Algorithm \ref{alg:partitioning} are summarized in Theorem \ref{thm:clustering}.
To prove the theorem, we need to show the properties of the exponential distribution Exp$(\beta)$.
By the cumulative distribution function of the exponential distribution, we have for any $x\geq 0$, the probability that $T\geq x$, denoted as $Pr[T\geq x]$, is $exp(-\beta x)$.
\[Pr[T\geq x]=exp(-\beta x).\]
The \emph{memoryless} property of the exponential distribution is that for any $s,t\geq 0$,
\[Pr[T\geq s+t\,|\,T\geq s]=Pr[T\geq t].\]

\begin{algorithm}[hbt]
	\caption{Cluster-Out(Cluster-In)$(G(V,E),U,r,s)$}
\label{alg:partitioning}
\begin{algorithmic}[1]
\renewcommand{\algorithmicrequire}{\textbf{Input:}}
\renewcommand{\algorithmicensure}{\textbf{Output:}}
\REQUIRE $G(V,E)$, a vertex set $U\subseteq V$, an integer $s\geq 2$ and a parameter $r>0$
\ENSURE $(V_1,\cdots,V_d)$
\vspace{1.3mm}

\STATE $\beta \leftarrow \log (s)/r$;
\STATE For each $u\in U$, picks a radius $r_u$ from Exp$(\beta)$ independently;
\STATE For each $v\in V$,  assigns $v$ to the cluster centered at vertex $u\in U$ which maximizes $r_u-d(u,v)$ if the quantity is positive; otherwise, do not assign $v$ to any cluster; (assigns $v$ to the cluster centered at vertex $u\in U$ which maximizes $r_u-d(v,u)$, respectively); \label{line:cluster_construct}
\STATE $(V_1,\cdots,V_{d-1}) \leftarrow $ the clusters constructed in Line \ref{line:cluster_construct};
\STATE $V_d\leftarrow V-\cup_{i\in [1,d-1]} V_i$;
\RETURN ($V_1,\cdots,V_{d-1},V_d$);
\end{algorithmic}
\end{algorithm}

\begin{theorem}
\label{thm:clustering}
For a graph $G(V,E)$, an arbitrary set of vertices $U\subseteq V$, a fixed but unknown set of vertices $S$ of size $s\geq 2$, and a parameter $r>0$, Algorithm 1 constructs a partitioning $(V_1,\cdots,V_d)$ of $V$ such that, \\
(1) all clusters $V_1,\cdots,V_{d-1}$, except for the last cluster $V_d$, have a radius at most $c\cdot r$ with probability at least $1-n/s^c$ for $c\geq 1$; \\
(2) the last cluster $V_d$ has size at most $|V|-|U|$; \\
(3) for every $u,v\in S \times V$ s.t. $d(u\rightleftarrows v)\leq R$, they are in the same cluster $V_i$ with probability at least $exp(-R/r\cdot \log s)$.\\
Furthermore, the algorithm runs in time $O(m\log n)$.
\end{theorem}

\proof
For brevity, we only prove the theorem for the routine Cluster-Out. The proof for Cluster-In is similar and omitted here.
We first prove property (1).
By construction, each cluster $V_i$ for $i\in [1,d-1]$ can have radius at most $r_u$, where $u$ is the center of $V_i$.
Then according to the cumulative distribution function of the exponential distribution, we have that 
\[Pr[r_u\geq c\cdot r]=exp(-c\cdot \beta r)\leq s^{-c}.\]
By union bound, the event that all clusters $V_1,\cdots,V_{d-1}$ have radius at most $c\cdot r$ happens with probability at least $1-n/s^c$.

Next we prove property (2).
We show that each $u\in U$ will be assigned to the cluster centered at $u$ itself, if there exists no vertex $u'\in U$ such that $r_{u'}-d(u',u)>r_{u}$. This is because $r_u-d(u,u)=r_u>0$ and it meets the requirements of adding a vertex to a cluster.  Otherwise, $u$ will be assigned to another cluster $u'\not=u$ which maximizes $r_{u'}-d(u',u)$.
In either case $u$ is included in some cluster $V_i$ for $i\in[1,d-1]$. That is,
\[|\cup_{i\in [1,d-1]} V_i|\geq |U|.\]
Then by construction we have
\begin{equation*}
\begin{split}
|V_d|=&|V-\cup_{i\in [1,d-1]} V_i| \\
\leq &|V|-|\cup_{i\in [1,d-1]} V_i| \\
\leq &|V|-|U|.
\end{split}
\end{equation*}

For property (3), we indeed prove a stronger argument: for every $u,v\in V \times V$, instead of $u,v\in S \times V$, at round-trip distance at most $R$, they are in the same cluster $V_i$ with probability at least $exp((-R/r)\log s)$.
Assume that $x\in U$ is the vertex which maximizes $r_x-\min\{d(x,u),d(x,v)\}$ and that the quantity is larger than $0$ (otherwise, both $u$ and $v$ are contained in $V_d$.)
Let $T$ be the second largest value of this quantity or zero, whichever is larger.
Suppose without loss of generality (w.l.o.g.) that $d(x,u)\leq d(x,v)$.
Then $u$ is assigned to the cluster $V_i$ centered at $x$ and $r_x-d(x,u)\geq T$.
$u$ and $v$ are separated in different balls only if $r_x-d(x,v)<T$.
This would imply that
\begin{equation*}
\begin{split}
&r_x-(d(x,u)+R)\\
\leq &r_x-d(x,u)-d(u\rightleftarrows v)\\
< &r_x-d(x,u)-d(u,v)\\
\leq &r_x-d(x,v)<T
\end{split}
\end{equation*}
The second to last inequality holds because of the triangle inequality.
By the memoryless property of the exponential distribution, the probability of both vertices $u$ and $v$ containing in the same cluster $V_i$ is at least
\[Pr[r_x\geq d(x,u)+R+T \,|\, r_x\geq d(x,u)+T] = exp(-\beta R) = exp(-R/r\cdot \log s).\]

We now analyze the running time. The algorithm can be implemented sequentially as follows. First, we add a virtual root $r$ and connect it to each of $u\in U$ with an edge of weight $0$. Then we perform a shortest path search from $r$ to all vertices in $V$. The partitioning of $V$ can be inferred easily from the shortest path tree. So the running time is dominated by that of a single-source shortest path search using Fibonacci heaps, $O(m+n\log n)=O(m\log n)$.
\qed


\subsection{Estimating the Sizes of In- and Out-Balls} \label{subsec:est}
Here we present a random sampling based algorithm for estimating the sizes of in- and out-balls around some selected vertices. It is an adaptation of Cohen's algorithm \cite{Cohen97} (see also a similar algorithm in \cite{PRS+18}) to the multiple source setting. The algorithm takes in a subset of vertices $U$ in a graph $G$ and parameters $r>0$, $\epsilon\in (0,1)$, and estimates the sizes $f_{In}^u$ and $f_{Out}^u$ of the $in$-$ball_{V}(u,r)$ and $out$-$ball_{V}(u,r)$ for every vertex $u$ in $U$ such that the estimations are no larger or smaller than $f_{In}^u$ and $f_{Out}^u$ by at most $\epsilon$, respectively. As in Algorithm \ref{alg:size}, we first sample a subset $T$ of vertices from $V$ and compute distances between vertices in $T$ and $U$, e.g., by constructing a single-source shortest path tree for each vertex in either $T$ or $U$, whichever has a smaller cardinality. Based on the distances, we can then calculate the estimations of $f_{In}^u$ and $f_{Out}^u$ with the desirable accuracy. The properties of Cohen's algorithm is summarzied in Lemma \ref{lem:est} while the properties of Algorithm \ref{alg:size} is summarized in Corollary \ref{cor:est}. In our subsequent algorithms, we will call Algorithm \ref{alg:size} with $U=V$. However, the algorithm enjoys the flexibility that can perform size estimations for some subset of vertices $U$, instead of all vertices $V$.

\begin{algorithm}[hbt]
	\caption{Estimate-In-Out-Balls($G(V,E),r,\epsilon,U=\{u_1,\cdots,u_d\}$)}
\label{alg:size}
\begin{algorithmic}[1]
\renewcommand{\algorithmicrequire}{\textbf{Input:}}
\renewcommand{\algorithmicensure}{\textbf{Output:}}
\REQUIRE G(V,E), a vertex set $U$, parameters $r>0$ and $\epsilon\in (0,1)$
\ENSURE $\widetilde{F}_{Out},\widetilde{F}_{In}$
\vspace{1.3mm}

\STATE Sample $t=\lceil 5\epsilon^{-2}\log n \rceil$ vertices $T=\{v_1,\cdots,v_t\}$ uniformly at random independently from $V$ with replacement;
\STATE Compute the distances between each of $U=\{u_1,\cdots,u_d\}$ and each of $T=\{v_1,\cdots,v_t\}$, e.g., by building a singe-source shortest path tree from each vertex in either $T$ or $U$, whichever has a smaller cardinality; \label{line:est:dist}
\STATE $\widetilde{F}_{Out} \leftarrow \{[\widetilde{f}_{Out}^{u_1},\cdots,\widetilde{f}_{Out}^{u_d}]\,|\,\widetilde{f}_{Out}^{u_i} \leftarrow$ the fraction of vertices $v\in T$ such that $d(u_i,v)\leq r$ for $i\in [1,d]\}$;
\STATE $\widetilde{F}_{In} \leftarrow \{[\widetilde{f}_{In}^{u_1},\cdots,\widetilde{f}_{In}^{u_d}]\,|\,\widetilde{f}_{In}^{u_i} \leftarrow$ the fraction of vertices $v\in T$ such that $d(v,u_i)\leq r$ for $i\in [1,d]\}$;
\RETURN ($\widetilde{F}_{Out},\widetilde{F}_{In}$);
\end{algorithmic}
\end{algorithm}

\begin{lemma}[\cite{Cohen97}] \label{lem:est}
For a graph $G(V,E)$, an arbitrary vertex set $U$ of size $d$, parameters $r>0$ and $\epsilon\in (0,1)$, let $F_{Out}$ ($F_{In}$) be the $n$-length vector such that the value corresponding to a vertex $u\in V$, $f_{Out}^u$ ($f_{In}^u$), is the fraction of vertices $v\in V$ such that $d(u,v)\leq r$ ($d(v,u)\leq r$, resp.). There is an algorithm that constructs their estimations $\widetilde{F}_{Out}$ and $\widetilde{F}_{In}$ such that for every $u\in V$, both the inequalities $\widetilde{f}_{Out}^u - f_{Out}^u \leq \epsilon$ and $\widetilde{f}_{In}^u - f_{In}^u \leq \epsilon$ hold w.h.p. Furthermore, the algorithm runs in time $O(m\epsilon^{-2}\log^2n)$.
\end{lemma}

\begin{corollary}
\label{cor:est}
For a graph $G(V,E)$, an arbitrary vertex set $U$ of size $d$, parameters $r>0$ and $\epsilon\in (0,1)$, let $F_{Out}$ ($F_{In}$) be the $d$-length vector such that the value corresponding to a vertex $u\in U$, $f_{Out}^u$ ($f_{In}^u$), is the fraction of vertices $v\in V$ such that $d(u,v)\leq r$ ($d(v,u)\leq r$, resp.). Algorithm 2 constructs their estimations $\widetilde{F}_{Out}$ and $\widetilde{F}_{In}$ such that for every $u\in U$, both the inequalities $\widetilde{f}_{Out}^u - f_{Out}^u \leq \epsilon$ and $\widetilde{f}_{In}^u - f_{In}^u \leq \epsilon$ hold w.h.p. Furthermore, the algorithm runs in time $O(\min\{\epsilon^{-2}\log n, d\}m\log n)$.
\end{corollary}

\proof
Under the Chernoff bound, we easily have that
\[Pr[\widetilde{f}_{Out}^u - f_{Out}^u > \epsilon] \leq 2\cdot exp(-2t\epsilon^2) \leq 2\cdot exp(-10\log n) = 2n^{-10}< 0.002.\]
By a similar application of the Chernoff bound, the inequality for $\widetilde{f}_{In}^u$ also holds.
The run-time is dominated by $O(\min\{t,d\})$ invocations of single-source shortest path search using Fibonacci heaps in Line \ref{line:est:dist}.
\qed

\subsection{Source-wise Round-trip Cover Construction} \label{subsec:cover}

With Algorithms 1 and 2 at hand, we are ready to describe the algorithm for constructing source-wise round-trip covers. As mentioned earlier, this algorithm is an adaptation of the algorithm of \cite{PRS+18} for constructing standard round-trip covers. Much of the adatation work is performed in a subroutine RecursiveCover for the simpler task of constructing a recursive cover that is a partitioning of the vertex set such that for every vertex pair of interest at bounded round-trip distance, they are in the same part with at least some fixed probability $p$. The main source-wise round-trip cover construction is then a union of sufficiently many recursive covers computed by RecursiveCover to raise the fixed probability $p$ to a high probability bound. Through careful adaptation of the exploration radius and the source vertex set in every recursion, we achieve a probability $p$ dependent on $\log s$ instead of $\log n$ required in the standard setting. That means only $\widetilde{O}(s^{1/k})$ recursive covers instead of $\widetilde{O}(n^{1/k})$ are sufficient to secure the high probability bound, leading to a smaller size of source-wise round-trip covers.

\begin{algorithm}[!hbt]
	\caption{RecursiveCover$(G(V,E),r,S)$}
\label{alg:cover}
\begin{algorithmic}[1]
\renewcommand{\algorithmicrequire}{\textbf{Input:}}
\renewcommand{\algorithmicensure}{\textbf{Output:}}
\REQUIRE $G(V,E)$, a source vertex set $S$, and a parameter $r>0$
\ENSURE $C$
\vspace{1.3mm}

\STATE Let $c\geq 1$ be a sufficiently large constant;
\IF{$V=\emptyset$ $||$ $S=\emptyset$}
	\RETURN $\emptyset$; \COMMENT{Handle the special cases when $V=\emptyset$ or $S=\emptyset$}
\ENDIF
\IF{$|S|=1$}
	\STATE Let $u$ be the only vertex in $S$; \COMMENT{Handle the special case when $|S|=1$}
	\STATE $B \leftarrow ball_{V}(u,r)$;
	\RETURN $C \leftarrow \{B\}$; \label{line:ball0}
\ENDIF
\STATE $(\widetilde{F}_{Out},\widetilde{F}_{In}) \leftarrow $ Estimate-In-Out-Balls$(G, c\cdot r, 1/8, V)$; \label{line:estimate}
\STATE Let $U_{Out} \leftarrow \{u\in V\,|\,\widetilde{f}_{Out}^u\geq 3/4\}$ and $U_{In} \leftarrow \{u\in V\,|\,\widetilde{f}_{In}^u\geq 3/4\}$;
\vspace{+0.1in}
\IF{$U_{Out}\cap U_{In}\not = \emptyset$} \label{line:cover_if1}
    \IF{$|U_{Out}\cap U_{In}|<1/4\cdot |V|$}
        \RETURN $\{V\}$; \label{line:cover_exit1} \COMMENT{Failure exit 1}
    \ENDIF
	\STATE Choose a vertex $u\in U_{Out}\cap U_{In}$ arbitrarily, and let the round-trip radius $r_u$ assigned to $u$ be a uniformly random value in $[2c\cdot r, 2(c+1)\cdot r]$;
	\STATE $B \leftarrow ball_{V}(u,r_u)$;
	\RETURN $C \leftarrow \{B\}$ $\cup$ RecursiveCover$(G(V-B),r,S-B)$; \label{line:ball}
\ELSE
	\IF{$|U_{Out}|\leq 1/2\cdot |V|$} \label{line:cover_if2}
		\STATE $(V_1,\cdots,V_d) \leftarrow$ Cluster-Out$(G,V-U_{Out},r,|S|)$; \label{line:clusterout}
	\ELSE
		\STATE $(V_1,\cdots,V_d) \leftarrow$ Cluster-In$(G,V-U_{In},r,|S|)$; \label{line:clusterin}
	\ENDIF
    \IF{$\max_{1\leq i\leq d}|V_i| > 7/8\cdot |V|$}
        \RETURN $\{V\}$; \label{line:cover_exit2} \COMMENT{Failure exit 2}
    \ENDIF
    \FOR{$1\leq i\leq d$}
        \STATE $S_i \leftarrow S\cap V_i$; \COMMENT{Update each source vertex set}
    \ENDFOR
    \RETURN $C \leftarrow \text{RecursiveCover}(G(V_1),r,S_1) \text{ } \cup \cdots \cup \text{ } \text{RecursiveCover}(\newline G(V_d),r,S_d)$; \label{line:case2}
\ENDIF
\end{algorithmic}
\end{algorithm}

The algorithm presented in Algorithm 3 takes in a source vertex set $S$ in a graph $G(V,E)$ and a parameter $r>0$. It begins by handling the special cases for the vertex sets $V$ and $S$. When $V=\emptyset$ or $S=\emptyset$, it directly returns an empty set. When $V\not=\emptyset$ and $S$ contains only one vertex $u$, it includes the ball $ball_V(u,r)$ of round-trip radius $r$ around $u$ into the returned cover $C$. Otherwise, it calls Estimate-In-Out-Balls (Algorithm 2) to get estimation for the sizes of the in-balls and out-balls around every vertex $v\in V$ with radius $c\cdot r$, where $c$ is a sufficiently large constant.
If there is a vertex $u$ that can reach and be reachable from many vertices, the algorithm directly includes its ball $B$ of round-trip radius randomly picked from $[2c\cdot r, 2(c+1)\cdot r]$ into the cover $C$, and then recurses on the remaining graph $G(V-B)$.
Otherwise, the algorithm applies Cluster-Out(-In) (Algorithm 1) to partition the graph into clusters and then recurses on each of the clusters.
By proving that for each recursion in either case above the graph vertex size is reduced by at least a constant factor, we have that there are at most logarithmic levels of recursions. Then by union bound we can reason the probability of vertex pairs of interest at bounded round-trip distance in the same ball over all recursions.

\begin{theorem}
\label{thm:cover}
For a graph $G(V,E)$, an $s$-sized source vertex set $S\subseteq V$ and a parameter $r>0$, Algorithm 3 constructs a collection $C$ of balls such that, \\
(1) for every $u\in S,v\in V$ at round-trip distance at most $R$, they are in the same ball of $C$ with probability at least $exp((-6R/r)\log n\log s)$; \\
(2) each ball in $C$ has round-trip radius $O(r)$ w.h.p; \\
(3) each vertex $v\in V$ is contained in at most one ball of $C$.  \\
Furthermore, the algorithm runs in time $O(m\log^3n)$.
\end{theorem}

\proof
We first define notations we will use in the proof. We will use $G$ and $V$ to denote the original graph and its vertex set respectively, and use $G(V')$ and $V'$ to denote the graph before each invocation of Algorithm 1 or 2 and its vertex set respectively. Obviously $V'\subseteq V$.
For a chosen sufficiently large constant $c\geq 1$, a single call of Algorithm 2 yields properties in Corollary \ref{cor:est} w.h.p. Similarly, a single call of Algorithm 1 yields property (1) in Theorem \ref{thm:clustering} w.h.p.
Therefore, by the union bound, the event that all calls of Algorithms 1 and 2 yield the properties in Theorem \ref{thm:clustering} and Corollary \ref{cor:est} happens w.h.p. and we assume this happens in the proof.

Conditioning on the above event, we prove that the algorithm never fails and terminates through the exit in Line \ref{line:cover_exit1} when the if-test in Line \ref{line:cover_if1} is true, or otherwise the exit in Line \ref{line:cover_exit2}. For the former case, let $u$ be such a vertex in $U_{Out}$ $\cap$ $U_{In}$. According to Corollary \ref{cor:est}, we have 
\[|out\text{-}ball_{V'}(u, c\cdot r)|\geq (3/4-1/8)\cdot |V'|=5/8\cdot |V'|.\]
Similarly, $|in$-$ball_{V'}(u, c\cdot r)|\geq 5/8\cdot |V'|$. Then the size of their intersection satisfies that
\[|out\text{-}ball_{V'}(u, cr) \cap in\text{-}ball_{V'}(u, cr)|\geq 1/4\cdot |V'|.\]
This implies that $|U_{Out}$ $\cap$ $U_{In}| \geq 1/4\cdot |V|$ and the failure in Line \ref{line:cover_exit1} would not occur.

For the latter case (when the if-test in Line \ref{line:cover_if1} fails), we have either $|U_{Out}|\leq 1/2\cdot |V'|$ or $|U_{In}|\leq 1/2\cdot |V'|$. Otherwise, their intersection must not be empty. We assume w.l.o.g. that $|U_{Out}|\leq 1/2\cdot |V'|$ and thus the algorithm Cluster-Out is performed (in Line \ref{line:clusterout}).
According to Corollary \ref{cor:est} and property (1) of Theorem \ref{thm:clustering}, we have for every $1\leq i\leq d-1$,
\[|V_i|\leq (3/4+1/8)\cdot |V'|=7/8\cdot |V'|.\]
Note that we use $V'-U_{Out}$ as the centering vertices when calling Cluster-Out. By property (2) of Theorem \ref{thm:clustering}, the last cluster $V_d$ has size at most 
\[|V_d|=|V'|-|V'-U_{Out}|\leq |U_{Out}|\leq 1/2\cdot |V'|.\]
Therefore, all clusters have size at most $7/8\cdot|V'|$ and the failure in Line \ref{line:cover_exit2} would not occur.

We now prove each of the three properties, staring from property (2). By the ball constructions in Lines \ref{line:ball0} and \ref{line:ball} and the fact that properties of Theorem \ref{thm:clustering} and Corollary \ref{cor:est} happen w.h.p., each ball has round-trip radius $O(r)$ in $G(V')$ w.h.p. Then each ball has round-trip radius $O(r)$ in $G$ w.h.p. because $V'$ is a subset of $V$.
Property (3) can be easily verified by construction.

Next we prove property (1). In the algorithm, only Lines \ref{line:ball} and \ref{line:case2} can separate two vertices into different balls. In Line \ref{line:case2}, the probability of not separating $u\in S,v\in V$ at round-trip distance at most $R$ in different balls, if they have not been separated before, is at least $exp(-R/r\cdot \log s)$, according to Theorem \ref{thm:clustering}. In Line \ref{line:ball}, the probability is at least 
\[1-R/2r\geq exp(-R/r\cdot \log n)\geq exp(-R/r\cdot \log s).\]
Note that in Lines  \ref{line:ball} and \ref{line:case2}, the sizes of $V'$ are multiplied by at most 7/8 and 3/8, respectively. Therefore, the total levels of recursions can be at most $\lceil\log_{7/8}|V|\rceil$. Then the probability of not separating $u$ and $v$ in different balls over all recursions is at least 
\[(exp(-R/r\log s))^{\lceil\log_{7/8}n\rceil}\geq exp((-6R/r)\log n\log s).\]

The running time $O(m\log^3n)$ follows because the total levels of recursions is $O(\log n)$ and the running time of each recursion is dominated by the running time of Estimate-In-Out-Balls (Algorithm 2) $O(m\epsilon^{-2}\log^2n)$.
\qed

The main algorithm for constructing source-wise round-trip covers is presented in Algorithm 4. Given a source vertex set $S$ in a graph $G$, a stretch parameter $k$ and a round-trip radius $R$, the union of the collection of balls obtained by $\widetilde{O}(s^{1/k})$ invocations of RecursiveCover (Algorithm 3) is an $S$-sourcewise round-trip cover of $G$ w.h.p. Executing RecursiveCover for $\widetilde{O}(s^{1/k})$ times suffices to ensure that the resulting collection of balls is the desired source-wise round-trip cover w.h.p.

\begin{algorithm}[hbt]
	\caption{SWRT-Cover$(G(V,E),k,R,S)$}
\label{alg:rtcover}
\begin{algorithmic}[1]
\renewcommand{\algorithmicrequire}{\textbf{Input:}}
\renewcommand{\algorithmicensure}{\textbf{Output:}}
\REQUIRE $G(V,E)$, a source vertex set $S$, an integer $k>1$, and a parameter $R>0$
\ENSURE An $S$-sourcewise $(O(k\log n),R)$-cover of $G$
\vspace{1.3mm}

\STATE $r \leftarrow 6Rk\log n$;
\STATE Let $c\geq 1$ be a sufficiently large constant;
\STATE $C \leftarrow \emptyset$;
\FOR{$1\leq i\leq c \cdot \lceil s^{1/k}\rceil  \cdot  \lceil\log n\rceil]$}
	\STATE $C \leftarrow C$ $\cap$ RecursiveCover$(G,r,S)$;
\ENDFOR
\RETURN $C$;
\end{algorithmic}
\end{algorithm}

\proof (\emph{Theorem \ref{thm:rtcover}.})
By property (1) of Theorem \ref{thm:cover}, we have that for every $u\in S,v\in V$ at round-trip distance at most $R$, the probability that they are in the same ball in a single invocation of RecursiveCover is at least 
\[exp((-6R/r)\log n\log s)=exp((-1/k)\log s)=s^{-1/k}.\]
The probability that $u$ and $v$ are in a ball $B\in C'$ for a structure $C'$ constructed by $\lceil s^{1/k}\rceil$ invocations of RecursiveCover is at least 
\[1-(1-1/s^{1/k})^{s^{1/k}}=1-exp(-1).\]
Then the probability that $u$ and $v$ are in a ball $B\in C$ for a structure $C$ constructed by $c \cdot \lceil s^{1/k}\rceil  \cdot  \lceil\log n\rceil$ invocations of RecursiveCover is at least
\[1-(exp(-1))^{c \cdot \lceil\log n\rceil}\geq 1-1/n^c.\]
For a sufficiently large constant $c$, the event that they are in a ball $B\in C$ happens w.h.p.

According to property (2) of Theorem \ref{thm:cover}, each ball $B\in C$ has round-trip radius at most $O(r)=O(Rk\log n)$ w.h.p.
Therefore, the constructed collection of balls $C$ is an $S$-sourcewise $(O(k\log n),R)$-cover w.h.p.
By property (3) of Theorem \ref{thm:cover}, each vertex is contained in $O(s^{1/k}\log n)$ balls of $C$.
\qed

\section{Removing the Dependence on the Edge Weights} \label{sec:dependence}
In this section, we remove the dependence on the edge weights for the bounds in Theorem \ref{thm:weightrts} in order to obtain our main theorem, Theorem \ref{thm:rts}.
Although there has been a technique \cite{PRS+18,EMP+16,RTZ08} that can remove the dependence on the edge weights for standard round-trip spanners, it is still not clear whether and how it can be adapted to the source-wise setting.
In this paper, we show an adaptation of the technique and rigorously prove its correctness.
Similar to \cite{PRS+18,EMP+16,RTZ08}, the intuition is that we do not need to consider all edges of the original graph when constructing a source-wise $(O(k\log n), R)$-roundtrip cover for a fixed round-trip radius $R$.
We introduce contraction operations in the source-wise case so that when exponentially increasing round-trip distances are considered, the number of times that each edge is presented is bounded by a logarithm in the number of vertices.

We first describe contraction operations for the standard setting as defined in \cite{PRS+18}, and then show the adapted contractions in the source-wise setting.
For a graph $G(V,E)$ and two real numbers $x_L$ and $x_R$, $G$ is contracted to $[x_L, x_R]$ by, (1) merging vertices in any strongly connected component (or for short \emph{SCC}) with maximum edge weight at most $x_L$ into a single vertex; (2) removing all edges of weights larger than $x_R$; (3) removing all edges that \emph{do not} participate in any \emph{SCC} with maximum edge weight at most $x_R$; (4) removing all vertices with no edge after the above steps.
In the source-wise setting, we have a source vertex set $S\subseteq V$ in the original graph $G$.
During the contraction, we can construct a new source vertex set $S'$ together with the contracted graph by adding additional work in some of the contraction steps.
Specifically, $S'$ is first initialized to the original sources $S$.
In Step (1), if a source vertex $u$ participates in an \emph{SCC} and gets merged into a single vertex $u'$, replace $u$ by $u'$ in $S'$. In Step (4), if a source vertex $u$ has no edge (after all the above steps) and gets removed, remove $u$ from $S'$.

We will use the following definition and lemma in our proof shortly.
\begin{definition}
(Definition 5.6 from \cite{PRS+18}) For two vertices $u$ and $v$ in a graph $G$, their $L_\infty$-roundtrip distance, $d_G^\infty(u,v)$, is defined as the minimum value of $d$ such that there is a (simple) cycle with maximum edge weight $d$ containing $u$ and $v$.
\end{definition}

\begin{lemma} \label{lem:contract}
(Lemma 5.7 from \cite{PRS+18}) For a graph $G$ and every $t\in \mathbb{Z}$, let $G^{(t)}$ be $G$ contracted to $[2^t/n, 2^t]$. The total number of edges and vertices in all non-empty $G^{(t)}$ are $O(m\log n)$ and $O(n\log n)$, respectively.
\end{lemma}

We will use the algorithm, Roundtrip-$L_\infty$-Spanner$(\cdot)$, in \cite{PRS+18} to construct, for a digraph $G$, a tree for efficient computations of $d_G^\infty(u,v)$ for each edge $(u,v)$ in $G$, and a sparse edge set preserving pair-wise $L_\infty$-roundtrip distances in $G$. The following lemma summarizes properties of the algorithm.

\begin{lemma} \label{lem:Linfty}
(Lemma 5.9 from \cite{PRS+18}) For a directed graph $G(V,E)$, Roundtrip-$L_\infty$-Spanner$(G)$ constructs, \\
(1) an $O(n)$-sized edge set $H\subseteq E$ such that for any vertex pair $u,v$ contained in a cycle with maximum edge weight $R$ in $G$, there is a cycle containing $u$ and $v$ with maximum edge weight $R$ in $G'(V,H)$; \\
(2) a tree $T$ such that for every $u,v\in V\times V$, the label of their lowest common ancestor in $T$ is $d_G^\infty(u,v)$. \\
Moreover, the algorithm runs in $O(m\log n)$ time.
\end{lemma}

\begin{corollary} \label{cor:contract}
Given the tree $T$ obtained in Lemma \ref{lem:Linfty}, all non-empty $G^{(t)}$ and their source vertex sets $S^{(t)}$ can be computed in linear time.
\end{corollary}

\proof
Consider an edge $e=(u,v)\in E$ that is an edge in $G^{(t)}$. Because $e$ is kept in $G^{(t)}$, we have $d_G^\infty(u,v)>2^t/n$. Also, $e$ must participate in an \emph{SCC} with maximum edge weight $2^t$ in $G$. By definition, it means $d_G^\infty(u,v)\leq 2^t$. Combining the two inequalities we have
\begin{equation} \label{eq:t}
\log_2 d_G^\infty(u,v)\leq t< \log_2 d_G^\infty(u,v)+\log n.
\end{equation}
It means that the edge $e$ is presented only in $G^{(t)}$ for $t$ in the above interval.

Consider a vertex $u$ presented in $G^{(t)}$. By construction, we have $u$ belongs to an \emph{SCC} with maximum edge weight $2^t$ in $G$. Then for $t'\geq t+\log_2 n$, $u$ will be merged into an \emph{SCC} in $G^{(t')}$. That is, $u$ only appears in at most $\log_2 n$ graphs $G^{(t)}$. 

Suppose initially $G^{(t)}=(V,\emptyset)$ for each $t\in \mathbb{Z}$. Given the tree $T$ computed in Lemma \ref{lem:Linfty}, we can calculate $d_G^\infty(u,v)$ for every edge $e=(u,v)$ by executing a lowest common ancester query between $u$ and $v$ in $T$ using constant time. We can then include $e$ into $G^{(t)}$ for every $\log_2 d_G^\infty(u,v)\leq t< \log_2 d_G^\infty(u,v)+\log n$ according to Inequality (\ref{eq:t}). Next, with the knowledge that each vertex $u$ can only appear in at most $\log_2 n$ graphs $G^{(t)}$, we merge vertices in an \emph{SCC} when necessary. Finally, all $G^{(t)}$ can be obtained by removing all vertices with no edge. Similarly, all contracted source vertex sets $S^{(t)}$ can also be calculated in the vertex merging and removing steps.  All these steps work in linear time. 
\qed

Now we are ready to present the algorithm (Algorithm \ref{alg:rtspanner}) for constructing source-wise round-trip spanners of size independent of the maximum edge weight. Consider a source vertex set $S\subseteq V$ in a digraph $G(V,E)$. We first call Roundtrip-$L_\infty$-Spanner on $G$ to construct an edge set $H_1$ and a tree $T$. Next, for every $t\in \mathbb{Z}$, we construct a contracted graph $G^{(t)}$ and its corresponding source vertex set $S^{(t)}$ with the help of $T$, as shown in Corollary \ref{cor:contract}. After that, for every non-empty $G^{(t)}$, we call SWRT-Cover (Algorithm \ref{alg:rtcover}) with parameters $k,2^t$ and $S^{(t)}$ to get a round-trip cover $C$. For each ball $B\in C$, we include edges in $RT$-$Tree(B)$ into the edge set $H$. Finally, $G'(V,H)$ is returned as the source-wise round-trip spanner.

\begin{algorithm}[hbt]
	\caption{SWRT-Spanner$(G(V,E),k,S)$}
\label{alg:rtspanner}
\begin{algorithmic}[1]
\renewcommand{\algorithmicrequire}{\textbf{Input:}}
\renewcommand{\algorithmicensure}{\textbf{Output:}}
\REQUIRE $G(V,E)$, a vertex set $S$, and an integer $k>1$
\ENSURE An $S$-sourcewise $O(k\log n)$-roundtrip spanner of $G$
\vspace{1.3mm}
\STATE $(H_1,T) \leftarrow $Roundtrip-$L_\infty$-Spanner$(G)$; \label{line:lemma3}
\STATE $H \leftarrow H_1$;
\STATE For every $t\in \mathbb{Z}$, construct $G^{(t)}$ and its corresponding source vertex set $S^{(t)}$ with the help of $T$;
\FOR{each non-empty $G^{(t)}$}
	\STATE $C \leftarrow $SWRT-Cover$(G^{(t)},k,2^t,S^{(t)})$;
	\STATE $H \leftarrow H\cup \cup_{B\in C}RT$-$Tree(B)$;
\ENDFOR
\RETURN $G'(V,H)$;
\end{algorithmic}
\end{algorithm}
\proof (\emph{Theorem \ref{thm:rts}})
According to Lemma \ref{lem:contract}, both numbers of edges and vertices in all non-empty $G^{(t)}$ are increased by a factor of $\log n$. Then according to Theorem \ref{thm:rtcover} and Corollary \ref{cor:contract}, the number of edges in the spanner and the running time are $O(ns^{1/k} \log^2 n)$ and $O(ms^{1/k} \log^5 n)$, respectively.

It remains to prove that the stretch factor is $O(k\log n)$. Consider vertices $u,v\in S\times V$ at round-trip distance $R$ in $G$ such that $2^{t-1}\leq R< 2^{t}$. Suppose vertices $u'$ and $v'$ are the corresponding vertex of $u$ and $v$ in the contracted graph $G^{(t)}$, respectively. By the definition of source-wise round-trip covers, in the structure $\cup_{B\in C}RT$-$Tree(B)$ for an $S$-sourcewise $(O(k\log n), 2^{t})$-cover $C$, there must exist a round-trip path $P'(u',v')$ of distance $O(k2^{t}\log n)$ between $u'$ and $v'$ in $G^{(t)}$. According to Lemma \ref{lem:Linfty}, for any \emph{SCC} with maximum edge weight $w'$ in $G$, there must exist a corresponding \emph{SCC} with maximum edge weight $w'$ in the graph $G_1(V,H_1)$, where $H_1$ is the sparse edge set constructed by Roundtrip-$L_\infty$-Spanner in Line \ref{line:lemma3}. Therefore, we can uncontract $P'(u',v')$ into a new round-trip path $P(u,v)$ in $G'(V,H)$, by first uncontracting every vertex resulted from vertex merging back to the vertices of its \emph{SCC} in $G$, and then adding edges of the corresponding \emph{SCC} in $H_1$ to $P$. This unpacking process enlarges the distance of $P'$ by at most 
\[w'\cdot n\leq 2^{t}/n\cdot n=2^t,\]
where the first inequality holds because $P'$ is a path in $G^{(t)}$. Therefore, the distance of $P$ is $O((k+1)2^{t}\log n)=O(kR\log n)$ as $2^{t-1}\leq R$. This complete the proof.
\qed

%

\section{Conclusion and Future Work} \label{sec:conclusion}
In this paper, we propose a fast algorithm for constructing source-wise round-trip spanners in weighted directed graphs.
Specifically, given an $n$-vertex $m$-edge graph $G(V,E)$, an $s$-sized source vertex set $S\subseteq V$ and an integer $k>1$.we propose a fast algorithm that in time $O(ms^{1/k}\log^5n)$ constructs an $S$-sourcewise round-trip spanner of stretch $O(k\log n)$ and size $O(ns^{1/k}\log^2n)$ w.h.p.
Essentially we propose a source-wise graph partitioning algorithm which takes an integer $s$ and a graph $G$ as input, partitions $G$ into clusters of bounded radius, and guarantees that for every $u,v\in S\times V$ at small round-trip distance, the probability of they are in different clusters is small.
We then rigorously prove that the recursive algorithm for constructing standard round-trip spanners \cite{PRS+18} can be generalized to the source-wise setting. Compared to the fast algorithms for constructing all-pairs round-trip spanners \cite{PRS+18,CLR+20}, our algorithm improves the running time and the size of the spanner when $k$ is super-constant. Compared with the existing algorithm for constructing source-wise round-trip spanners \cite{ZL17}, the developed algorithm significantly improves their construction time $\Omega(\min\{ms,n^\omega\})$ to nearly linear $O(ms^{1/k}\log^5n)$ at the expense of an extra factor $\log n$ in the stretch.
As the future work, we will study deterministic constructions of source-wise round-trip spanners and how to improve the $\log n$ factor in the stretch while preserving the nearly linear running time. Note that such improvements in the stretch have been achieved for all-pairs spanners \cite{CLR+20}. It is also interesting to study general \emph{pair-wise} round-trip spanners and investigate whether efficient algorithms exist when there is no structure in the vertex pairs, e.g., $S\times V$ in the source-wise case.

\bibliographystyle{plain}
\bibliography{fass}

\end{document}